\documentclass[twocolumn,showpacs]{revtex4}

\usepackage{graphicx}
\usepackage{dcolumn}
\usepackage{amsthm}
\usepackage{amsmath}

\begin{document}

\preprint{}
\title{CPT Symmetry and The Equality of Mass and Lifetime}

\author{J.C. Yoon}
\email{jcyoon@u.washington.edu} \affiliation{University of
Washington}

\date{\today}

\begin{abstract}
$CPT$ theorem has been known to imply the equality of mass and
lifetime between particle and antiparticle even if $C$(charge
conjugation) symmetry is violated. However, its mathematical
verification is insufficient and limited as it considers only one
type of $C$ violation. Here $C$, not $CPT$, symmetry will be shown
to imply the equality of mass and coupling constants and the
lifetime could be the same regardless of $C$ and $CPT$ violation.
And we conclude that the equality of mass and lifetime is
prerequisite for the $CPT$ theorem to be valid and it is not
implied by $CPT$ symmetry.
\end{abstract}

\maketitle

\section{Introduction}
It is commonly believed that the $CPT$ theorem implies that a
particle and its antiparticle must have the same mass and
lifetime\cite{Murayama}. However, no rigorous theoretical
investigation has been provided especially when $C$ symmetry is
violated. Let us first review and define $C$ symmetry and particle
and antiparticle symmetry with care and investigate the proofs of
the mass and lifetime equality.

\section{The Definition of Symmetry}

A particle and its antiparticle can be described by two
independent Dirac equations
\begin{eqnarray}
(i \partial\!\!\!\slash  - m)\psi &=& + e A_{\mu}\gamma^{\mu}\psi
\qquad \
\mathrm{for~particle} \nonumber \\
(i \partial\!\!\!\slash  - \overline{m})\psi^{c} &=& -
\overline{e} A^{c}_{\mu}\gamma^{\mu} \psi^{c} \qquad
\mathrm{for~antiparticle} \nonumber
\end{eqnarray}
and they are expected to obey the same equation with the same mass
($\overline{m}=m$)and opposite
charges($\overline{e}=e$)\cite{ItzyksonZuber}.
Under the transformation of particle and antiparticle on $\psi$
and the potential $A$,
\begin{eqnarray}
\psi \rightarrow \psi^{c}=\eta_{c}C{\overline{\psi}}^{T} \nonumber
\\
A_{\mu} \rightarrow A_{\mu}^{c} = - A_{\mu}. \nonumber
\end{eqnarray}
we have
\begin{eqnarray}
(i \partial\!\!\!\slash  - m)\psi^{c} &=& - e
A^{c}_{\mu}\gamma^{\mu}\psi^{c}  \nonumber \\
(i \partial\!\!\!\slash  - \overline{m})\psi^{c} &=& -
\overline{e} A^{c}_{\mu}\gamma^{\mu} \psi^{c} \qquad
\mathrm{for~antiparticle} \nonumber
\end{eqnarray}
However, these charge conjugation transformations of $\psi
\rightarrow \psi^{c}$, $A_{\mu} \rightarrow A_{\mu}^{c}$ is
insufficient and it still requires the same mass and coupling
constant($\overline{m}=m$ and $\overline{e}=e$) to obtain the
equation for antiparticle.

Let us define $charge~conjugation(C)$ as a symmetry that we can
find between the equations for particle and antiparticle and
$\it{particle~and~antiparticle~symmetry}$ by the equality of mass
and lifetime. $C$ symmetry could be violated when an antiparticle,
for example, simply have different mass or coupling constants in
strength, i.e. $m \neq \overline{m}$ or  $e \neq \overline{e}$,
\begin{eqnarray}
(i \partial\!\!\!\slash  - m)\psi &=& + e A_{\mu}\gamma^{\mu}\psi
\nonumber \\
(i \partial\!\!\!\slash  - \overline{m})\psi^{c} &=& -
\overline{e} A^{c}_{\mu}\gamma^{\mu}\psi^{c} \nonumber
\end{eqnarray}
and particle and antiparticle symmetry is also violated as they
have different masses and lifetimes. Another type of $C$ violation
could be found when more than two interactions involve. Let us
introduce the interactions with coupling constants $g_{1}$,
$g_{2}$ and a potential that violates $C$ symmetry as they have
different signs from $e A_{\mu}\gamma^{\mu}\psi$ for antiparticle,
\begin{eqnarray}
(i \partial\!\!\!\slash  - m)\psi &=& + e A_{\mu}\gamma^{\mu}\psi
+ g_{1} A_{\mu}\Gamma^{\mu} \psi + g_{2} B_{\mu}\Gamma^{\mu} \psi
\nonumber \\
(i \partial\!\!\!\slash  - m)\psi^{c} &=& - e
A^{c}_{\mu}\gamma^{\mu}\psi^{c} + g_{1} A^{c}_{\mu}\Gamma^{\mu}
\psi^{c} + g_{2} B^{c}_{\mu}\Gamma^{\mu} \psi^{c} \nonumber
\end{eqnarray}
where $\Gamma_{\mu}$ represents the general structure of
interaction. The lifetime $\tau$ of the particle is given by the
reciprocal of the total decay rate($\tau = 1 / \Gamma$)
\begin{eqnarray}
\Gamma &\propto&  |\mathcal{M}_{e}|^{2} + |\mathcal{M}_{g1} |^{2}
+ |\mathcal{M}_{g2} |^{2} \nonumber
\end{eqnarray}
if the final particles are different and
\begin{eqnarray}
\Gamma &\propto& |\mathcal{M}_{e} \pm \mathcal{M}_{g1} |^{2} +
|\mathcal{M}_{g2} |^{2} \nonumber
\end{eqnarray}
if we have the interference of $\mathcal{M}_{e}$ and
$\mathcal{M}_{g1}$ interactions decaying into the same final
particles. In general, we have
\begin{eqnarray}
\Gamma &\propto& |\mathcal{M}_{e}e^{i\phi_{e\pm}} \pm
\mathcal{M}_{g1}e^{i\phi_{g\pm}} |^{2} + |\mathcal{M}_{g2} |^{2}
\nonumber
\end{eqnarray}
where $\phi_{e\pm}$, $\phi_{g\pm}$ four independent phases for
particle and antiparticle. We may have different lifetimes for
particle and antiparticle due to the possible interferences, but,
the relative signs of interference and interaction terms in the
Dirac equation may not be explicitly related due to phase factors.
Therefore, $C$ symmetry could be violated even if a particle and
its antiparticle have the same mass and lifetime and the lifetimes
of particle and antiparticle could be different even when $C$
symmetry is conserved as the interference of interactions occurs.
It is because the equations based on for the definition of $C$
symmetry are not fundamental to the practical calculation of
interactions, lifetime. When we define $C$ symmetry, the
interactions are considered as a part of the equations,
\begin{eqnarray}
(i \partial\!\!\!\slash  - m)\psi &=& \pm e
A_{\mu}\gamma^{\mu}\psi \qquad \ \mathrm{for~defining~C} \nonumber
\end{eqnarray}
but the practical calculation of lifetime is based on the free
Dirac equation and interactions in separate, which neglects their
relative signs
\begin{eqnarray}
(i \partial\!\!\!\slash  - m)\psi &=& 0 \qquad \mathrm{and} \quad
\pm e A_{\mu}\gamma^{\mu} \psi \qquad \mathrm{for~lifetime}
\nonumber
\end{eqnarray}
and the opposite signs are known from other independent physical
observations indicating two opposite charges.

\section{Proof of mass equality }

Let us review how the mass and lifetime equality between particle
and antiparticle are proved and investigate these
proofs\cite{TDLee,LeeYang57}.

\newtheorem*{Theo}{Theorem}
\begin{Theo}
Mass equality between particles and antiparticles
\end{Theo}

\begin{proof}
Let $| p \rangle_{m}$ be the state of an elementary particle at
rest with its $z$ component angular momentum $m$, Since the mass
of the particle is given by the expectation value,
\begin{eqnarray}
mass_{p} = \langle p | H | p \rangle_{m} \label{eq:mass} \nonumber
\end{eqnarray}
where $H$ is the total Hamiltonian, real and independent of $m$.
Hence it equals its complex conjugation. We have
\begin{eqnarray}
mass_{p} = \langle p | H | p \rangle^{*}_{m} = \langle p |
\it{\Theta}^{-1} \it{\Theta} H \it{\Theta}^{-1} \it{\Theta} | p
\rangle_{m}. \nonumber
\end{eqnarray}
where the operator $\it{\Theta}$ is defined to be
\begin{eqnarray}
\it{\Theta} \equiv C P T. \nonumber
\end{eqnarray}
Apart from a multiplicative phase factor, under $C$ the state
becomes $| \overline{p} \rangle_{m}$, under $P$ it remains itself,
but under $T$, m is changed into $-m$, as given by
\begin{eqnarray}
T | j, m \rangle &=& U_{T} | j, m \rangle^{*} = U_{T} | j, m
\rangle = e^{i \pi J_{y}}| j, m \rangle \nonumber \\
&=& (-1)^{j+m} | j, -m \rangle \nonumber
\end{eqnarray}
Therefore,
\begin{eqnarray}
\it{\Theta} | p \rangle_{m} = e^{i\theta} | \overline{p}
\rangle_{-m} \nonumber
\end{eqnarray}
Since $\it{\Theta} H \it{\Theta}^{-1} = H$ by the CPT theorem, the
above expression can also be written as
\begin{eqnarray}
mass_{p} &=& \langle \overline{p} | H | \overline{p}
\rangle_{-m} \nonumber \\
mass_{\overline{p}} &\equiv& \langle \overline{p} | \overline{H} |
\overline{p} \rangle_{-m} \nonumber
\end{eqnarray}
Therefore,
\begin{eqnarray}
mass_{p} &=& mass_{\overline{p}}  \nonumber
\end{eqnarray}
when $H = \overline{H}$.
\end{proof}

In this proof, not only do we need the CPT theorem ($\it{\Theta} H
\it{\Theta}^{-1} = H$), but also it requires that $H =
\overline{H} \equiv C H C^{-1}$. The operation of CPT on a
particle already implies $C$ symmetry($H = \overline{H})$. If the
particle remains itself under $P$, the Hamiltonian should be also
be the same
\begin{eqnarray}
H = PHP^{-1} \nonumber
\end{eqnarray}
and since the mass of a particle is the same regardless of $z$
component of angular momentum, $m$ can be ignored and thus
\begin{eqnarray}
H_{mass} = TH_{mass}T^{-1}   \qquad   \mathrm{for~the~mass~term}
\nonumber
\end{eqnarray}
Now we have
\begin{eqnarray}
\it{\Theta} H_{mass} \it{\Theta}^{-1} &=&
CPTH_{mass}T^{-1}P^{-1}C^{-1} \nonumber \\
&=& CPH_{mass}P^{-1}C^{-1} \nonumber \\
&=& CH_{mass}C^{-1} \nonumber \\
&=& H_{mass}    \nonumber
\end{eqnarray}
by definition $\overline{H} \equiv C H C^{-1}$ and thus $H =
\overline{H}$.

However, this proof would fail with $C$ violation even if $CPT$
symmetry is conserved. For example, if CPT theorem
holds($\it{\Theta} H \it{\Theta}^{-1} = H$) but C is violated in a
way $\overline{H} \equiv C H C^{-1} = -H$, then the mass of
particle and antiparticle is different.
\begin{eqnarray}
mass_{p} &=& \langle p | H | p \rangle_{m} \nonumber \\
 &=& - \langle p |C^{-1}C H C^{-1} C| p \rangle_{m} \nonumber \\
&=& - \langle \overline{p} | \overline{H} | \overline{p} \rangle_{m} \nonumber \\
&\equiv& - mass_{\overline{p}} \nonumber
\end{eqnarray}
And in order to investigate whether the masses of particle and
antiparticle, the equation of particle and antiparticle is of
interest, not that of CPT transformed particle is unless it is
identified as antiparticle($\it{\Theta} H \it{\Theta}^{-1} = C H
C^{-1}$).
\begin{eqnarray} (i \partial\!\!\!\slash  - m)\psi &=&
0\qquad
\mathrm{for~particle} \nonumber \\
(i \partial\!\!\!\slash  - m)\psi^{CPT} &=& 0 \qquad
\mathrm{for~CPT~transformed~particle} \nonumber \\
(i \partial\!\!\!\slash  - \overline{m})\psi^{c} &=& 0\qquad
\mathrm{for~antiparticle}  \nonumber
\end{eqnarray}
Therefore, the mass of particle and antiparticle should be proved
by the conservation of $C$($H = \overline{H} \equiv C H C^{-1}$),
not by the $CPT$ theorem, as in
\begin{eqnarray}
mass_{p} &=& \langle p | H | p \rangle_{m} \nonumber \\
 &=& \langle p |C^{-1}C H C^{-1} C| p \rangle_{m} \nonumber \\
&=& \langle \overline{p} | \overline{H} | \overline{p} \rangle_{m} \nonumber \\
&\equiv& mass_{\overline{p}} \nonumber
\end{eqnarray}

To be more accurate, the equality of mass is prerequisite for $C$
symmetry since the transformation of equation($C H C^{-1}$) cannot
be equated to $\overline{H}$ unless its mass and coupling
constants consisting of the equation are the same. The free Dirac
equation(or the Dirac Hamiltonian), for example, cannot be
transform from one to another($\mathcal{L}_{0} \rightarrow
\overline{\mathcal{L}}_{0}$) unless $m = \overline{m}$.
\begin{eqnarray}
\mathcal{L}_{0} &=& \overline{\psi}(i\gamma^{\mu}\partial_{\mu} -
m)
\psi \nonumber \\
\overline{\mathcal{L}}_{0} &=&
\overline{\psi}(i\gamma^{\mu}\partial_{\mu} - \overline{m}) \psi
\nonumber
\end{eqnarray}
Therefore, the mass of particle and antiparticle is required to be
the same for the conservation of $C$ symmetry ($H = \overline{H}
\equiv C H C^{-1}$) and the proper definition of mass
representation should show that $C$ symmetry, not the $CPT$
theorem, implies the mass equality.

\section{Proof of lifetime equality }
Consider a Hamiltonian
\begin{eqnarray}
H = H_{strong} + H_{weak} \nonumber
\end{eqnarray}
where both terms are invariant under a proper Lorentz
transformation and $H_{strong}$ is assumed to be invariant under
$C$,$P$(Parity), and $T$(Time reversal), for example,
\begin{eqnarray}
CH_{strong}C^{-1} = H_{strong} \nonumber
\end{eqnarray}

\begin{Theo}
If a particle $A$ decays through the interaction $H_{weak}$, and
if the particle and its antiparticle $\overline{A}$ do not decay
into the same final products (as e.g. when $A$ is charged), then
to the lowest order of $H_{weak}$ the lifetimes of $A$ and
$\overline{A}$ are the same, even if $H_{weak}$ is not invariant
under charge conjugation.
\end{Theo}

\begin{proof}
Consider particle $A$ with spin zero and the final states $B$ and
$\overline{B}$ in the decays would also have spin zero.
\begin{eqnarray}
A \rightarrow B, \quad \overline{A} \rightarrow \overline{B},
\nonumber
\end{eqnarray}
Using the identity
\begin{eqnarray}
\langle \psi_{1} | \psi_{2} \rangle^{*} = \langle T\psi_{1} | T
\psi_{2} \rangle,  \nonumber
\end{eqnarray}
one obtains
\begin{eqnarray}
\langle B | H_{weak} | A \rangle^{*} &=& \langle TB |
TH_{weak}T^{-1}| TA \rangle \nonumber \\
&=& \langle TB | C^{-1}P^{-1}H_{weak}PC| TA \rangle, \nonumber
\end{eqnarray}
by the CPT theorem. If $H_{weak}$ commutes (or anticommutes) with
$P$, then
\begin{eqnarray}
\langle B | H_{weak} | A \rangle^{*} = \pm \langle TB |
C^{-1}H_{weak} C | TA \rangle. \nonumber
\end{eqnarray}
For a spinless system,
\begin{eqnarray}
|TA \rangle = | A \rangle, | TB \rangle = | B \rangle. \nonumber
\end{eqnarray}
Hence
\begin{eqnarray}
\langle B | H_{weak} | A \rangle^{*} &=& \pm \langle B| C^{-1}
H_{weak} C | A \rangle \nonumber \\
&=& \pm \langle CB | H_{weak} | CA \rangle = \pm \langle
\overline{B} | H_{weak} | \overline{A} \rangle. \nonumber
\end{eqnarray}
This shows that the lifetimes of $A$ and $\overline{A}$ are the
same.
\end{proof}
Following this proof, one can also show the lifetime equality of
particle and antiparticle under $CPT$ violation.
\newtheorem*{Theo2}{Theorem}
\begin{Theo2}
The lifetimes of $A$ and $\overline{A}$ are the same, even if
$H_{weak}$ is not invariant under CPT.
\end{Theo2}
\begin{proof}
The Hamiltonian $H$ commutes with $\it{\Theta} \equiv CPT$ by the
CPT theorem
\begin{eqnarray}
 \it{\Theta} H \it{\Theta}^{-1} = + H \nonumber
\end{eqnarray}
and if CPT is violated, then we have
\begin{eqnarray}
 \it{\Theta} H \it{\Theta}^{-1} = - H   \nonumber
\end{eqnarray}

Following the same steps as before, one obtains
\begin{eqnarray}
\langle B | H_{weak} | A \rangle^{*}
&=& - \langle TB | C^{-1}P^{-1}H_{weak}PC| TA \rangle, \nonumber
\end{eqnarray}
if CPT is violated(
$C^{-1}P^{-1}T^{-1} H_{weak} T P C = - H_{weak}$ ). If $H_{weak}$
commutes (or anticommutes) with $P$, then
\begin{eqnarray}
\langle B | H_{weak} | A \rangle^{*} = \mp \langle TB |
C^{-1}H_{weak} C | TA \rangle.  \nonumber
\end{eqnarray}
For a spinless system,
\begin{eqnarray}
\langle B | H_{weak} | A \rangle^{*}
&=& \mp \langle \overline{B} | H_{weak} | \overline{A} \rangle.
\nonumber
\end{eqnarray}
This shows that the lifetimes of $A$ and $\overline{A}$ are the
same.
\end{proof}

These theorems are limited as they exclude $C$ violations of
different mass and coupling constants($m \neq \overline{m}$ or $e
\neq \overline{e}$) where we cannot find a proper transformation
of equations($\overline{H} \neq C H C^{-1}$ and thus
$|\overline{A} \rangle \neq C |A \rangle$). If $C$ symmetry is
violated with different coupling constants, then the lifetimes of
particle and antiparticle would be different. However, if particle
and antiparticle have the same mass and coupling constants, the
lifetimes of particle and antiparticle are the same regardless of
$C$ and $CPT$ symmetry if no interference is assumed
\begin{eqnarray}
\Gamma &\propto&  |\mathcal{M}_{w}|^{2} + |\mathcal{M}_{g} |^{2}
\nonumber
\end{eqnarray}
where the signs of S matrix $\mathcal{M}_{w}$, $\mathcal{M}_{g}$
are irrelevant to the physical observation of lifetime.

\section{Conclusion}

The definitions and differences of $C$ symmetry and particle and
antiparticle symmetry are discussed and the proofs of mass and
lifetime equality are reviewed and criticized for their ambiguity
and exclusiveness. The conservation of $C$, not $CPT$, symmetry
requires the mass and coupling constants to be the same between
particle and antiparticle and the lifetime could be the same
regardless of $C$ and $CPT$ violation. Therefore, we can conclude
that the $CPT$ theorem does not imply the equality of mass and
lifetime and $\it{particle~and~antiparticle~symmetry}$, not $CPT$,
is appropriate when the equality of mass and lifetime is implied
\cite{Murayama,McKeownVogel,BGP,BMW,Kam,KTeV,BBLY}.

\end{document}